\def\year{2022}\relax
\documentclass[letterpaper]{article} 
\usepackage{aaai22}  
\usepackage{times}  
\usepackage{helvet}  
\usepackage{courier}  
\usepackage[hyphens]{url}  
\usepackage{graphicx} 
\usepackage{natbib}  
\usepackage{caption} 
\DeclareCaptionStyle{ruled}{labelfont=normalfont,labelsep=colon,strut=off} 
\usepackage{algorithm}
\usepackage{algorithmic}

\usepackage{newfloat}
\usepackage{listings}

\usepackage{booktabs}
\usepackage{amsmath}
\usepackage{bbding}
\usepackage{multirow}

\floatstyle{ruled}
\newfloat{listing}{tb}{lst}{}
\floatname{listing}{Listing}
\title{DoubleH: Twitter User Stance Detection via Bipartite Graph Neural Networks}
\author {
    Chong Zhang\textsuperscript{\rm 1},
    Zhenkun Zhou\textsuperscript{\rm 2}\thanks{Correspondence to: Zhenkun Zhou.},
    Xingyu Peng\textsuperscript{\rm 1},
    Ke Xu\textsuperscript{\rm 1}
}
\affiliations{
    \textsuperscript{\rm 1} State Key Lab of Software Development Environment, Beihang University, Beijing, China\\
    \textsuperscript{\rm 2} School of Statistics, Capital University of Economics and Business, Beijing, China\\
    chongzh@buaa.edu.cn, zhenkun@cueb.edu.cn, xypeng@buaa.edu.cn, kexu@buaa.edu.cn
}

\begin{document}

\maketitle

\begin{abstract}
Given the development and abundance of social media, studying the stance of social media users is a challenging and pressing issue. Social media users express their stance by posting tweets and retweeting. Therefore, the homogeneous relationship between users and the heterogeneous relationship between users and tweets are relevant for the stance detection task. Recently, graph neural networks (GNNs) have developed rapidly and have been applied to social media research. In this paper, we crawl a large-scale dataset of the 2020 US presidential election and automatically label all users by manually tagged hashtags. Subsequently, we propose a bipartite graph neural network model, DoubleH, which aims to better utilize homogeneous and heterogeneous information in user stance detection tasks. Specifically, we first construct a bipartite graph based on posting and retweeting relations for two kinds of nodes, including users and tweets. We then iteratively update the node's representation by extracting and separately processing heterogeneous and homogeneous information in the node's neighbors. Finally, the representations of user nodes are used for user stance classification. Experimental results show that DoubleH outperforms the state-of-the-art methods on popular benchmarks. Further analysis illustrates the model's utilization of information and demonstrates stability and efficiency at different numbers of layers.
\end{abstract}

\section{Introduction}
Stance detection is the task of automatically mining text author's position from text~\cite{SDdef}, and has been widely applied in political elections, public opinion research, and other fields. Social media provides a platform for people to express their opinions. Studies on the 2016 US presidential election~\cite{2016ele,2016ele2} and the 2019 Argentine presidential election~\cite{2019ele} indicate the research value of social media data for the election event. In recent years, with the increase in the number of people online on social media, Social media is more closely associated with political events~\cite{background}. Therefore, how to analyze this information accurately and efficiently is an urgent problem.

In social media networks, users and texts are the main targets of stance detection. In the user stance detection task, the stances of users are obtained by user-related information. Supervised methods exploit varied features such as tweets, user profile information, following, retweeting, and other features to train the classifier~\cite{su1,su2,su3}. In social networks, members of homogeneous groups tend to share similar stances~\cite{homoid, homoid2} on various topics. The idea of homogeneity is widely used in semi-supervised and unsupervised methods. The semi-supervised methods are grounded on the labels of some users, and employ label propagation on the user's following network~\cite{sesu1} or retweet network~\cite{sesu2,sesu3}, to label other users. The unsupervised methods utilize user similarity information to label users through user network~\cite{retweet-bert} or clustering~\cite{unsu1}. 

Most of the existing research on user stance detection is devoted to the feature selection and combination of text and network information. Unsupervised or semi-supervised methods find similar user pairs through the relationship between users and train models based on user information. However, it is difficult for these methods to outperform supervised learning methods. Manually labeling is time-consuming, and the user's stance can be better obtained if more labeled data and higher labeling efficiency are available. On the other hand, existing methods consider the characteristics of users such as user profiles and tweets~\cite{info1,info2}, or pay more attention to the relationship between users~\cite{rela1,retweet-bert,unsu1}. These methods interact between users and other users with homogeneous information. However, users are not the only objects that users interact with. For retweet relations between users, users establish a retweet relation through tweets as a bridge. The tweet is a text with rich semantic information and forms a binary relationship with users as heterogeneous information. Therefore, we aim to label user data automatically and better utilize and balance homogeneous and heterogeneous information.

In this paper, we focus on the 2020 US presidential election. We first collect all relevant users and tweets before the election and divide users into supporting either Joe Biden or Donald Trump. For massive amounts of user data, manual labeling methods such as crowdsourcing or preset political user lists~\cite{gcnso1} are not suitable. We notice that the hashtags in tweets represent the topics the tweets are about. In election events, the hashtag is an expression of user sentiment or opinion with a simple structure and clear tendency. The effectiveness of hashtags in stance labeling has been indicated in previous studies~\cite{label_bovet}. Therefore, we manually collect and label a collection of hashtags in which each hashtag is tagged as supporting Biden or Trump. Then the tweets are labeled according to the hashtags in the tweets. Tweets are an expression of user opinion, so we calculate the label ratio of each user's tweets and set a threshold according to the distribution of the tweet label ratio of the user. Finally, we annotate all users by comparing the tweet label ratio with the threshold.

\begin{figure*}[!htb]
    \centering
    \includegraphics[width=0.95\textwidth]{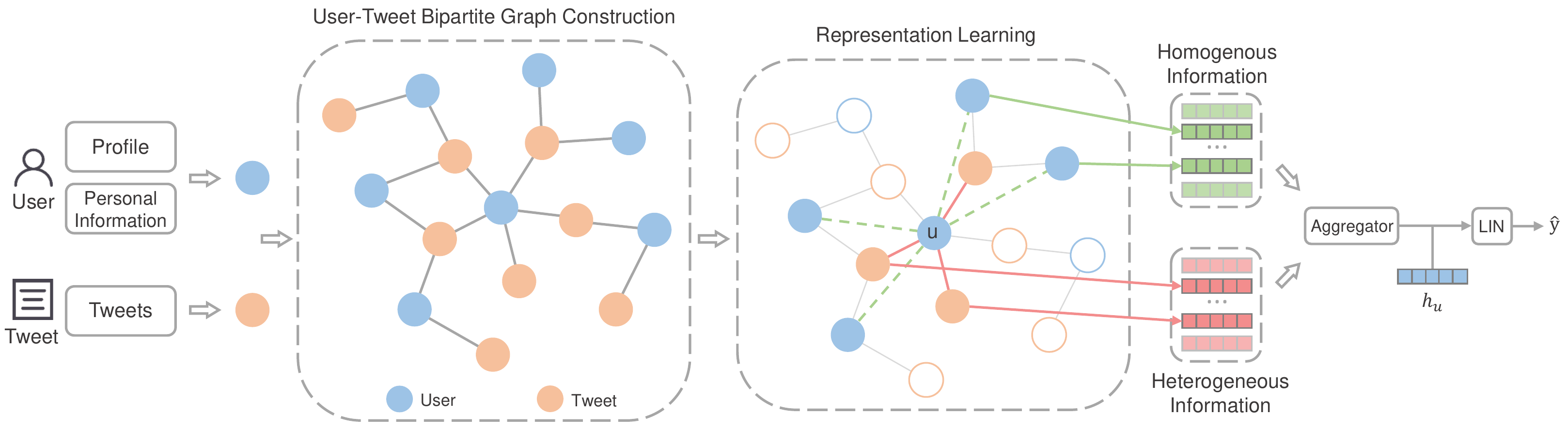}
    \caption{The framework of DoubleH. $u$ represents the target node, $h_u$ is the final representation of $u$.}
    \label{fig:model}
\end{figure*}

In recent years, graph neural networks (GNNs) have developed swiftly and have state-of-the-art performance in multiple fields. GNNs learn the representation of each node by aggregating the information of neighboring nodes and can retain structural information in the graph embedding. In the user's stance detection task, GNNs extract and mine the information in the retweet graph of users and tweets, and learn the representation of nodes grounded on the relationship between nodes. In the information aggregation of GNNs, user and tweet nodes can be regarded as information of different types of nodes or can be considered as text nodes and employ homogeneous graph neural networks to learn the representations. In this paper, we propose a novel graph neural network based on a user-tweet bipartite graph for user stance detection, named DoubleH. DoubleH aims to exploit both \textbf{H}omogeneous and \textbf{H}eterogeneous information to update node representation. As shown in Figure~\ref{fig:model}, users and tweets are used as two different types of nodes to construct the user-tweet bipartite graph. In each iteration, the heterogeneous information represented by one-hop neighbors and the homogeneous information of two-hop neighbors are extracted and processed separately, and then jointly interact with the central node. DoubleH leverages the learned node representations to classify user stances. So far, we make better use of homogeneous and heterogeneous information. Several experiments are conducted to verify the advantages of our method over the baselines.
To sum up, our contributions are as follows:
\begin{itemize}
\item We collect users and tweets related to the 2020 US presidential election, automatically label all user data based on hashtags, and create a large-scale dataset for supervised learning.
\item We propose a new method DoubleH, which takes the user's profile and personal information and tweets as initial information, aiming to parse and aggregate homogeneous and heterogeneous information of the user-tweet bipartite graph.
\item The experimental results show that DoubleH not only outperforms several graph-based baselines but also possesses decent efficiency. Further experiments illustrate the effect of the number of layers on the model and the different utilization of homogeneous and heterogeneous information.
\end{itemize}

\section{Related Work}
We classify existing user stance detection research into supervised learning, semi-supervised and unsupervised learning according to whether fully labeled data is required. In addition, graph neural networks have achieved good performance in social network research, we finally introduce methods related to graph neural networks and other advanced GNNs.

\subsection{Stance Detection}
\paragraph{Supervised Methods.}
Supervised stance detection models exploit labeled data to train a classifier to assign stance labels. Some content-based methods train classifiers with textual information such as tweets, user profiles, and hashtags. Hashtags in tweets were used as an important feature to classify users’ political positions in an earlier study~\cite{hashtag}. Socio-linguistic features such as utterances, e.g., emoticons, abbreviations, and word n-grams were used to distinguish between Republicans and Democrats~\cite{info2}. Some deep learning models for encoding text have also been applied to stance detection tasks, \citet{lstm} employed LSTM and Bidirectional LSTM to encode the tweet text on stance detection. \citet{bert} extracted contextual word embeddings of tweets by the large-scale pre-trained language model and applied a convolutional neural network (CNN) for further feature extraction for training a stance detection model. However, the application of advanced deep learning models is at the tweet level, not the user level. With the emergence of homophily ideology~\cite{homoid,homoid2}, it is observed that users interact more with people who share similar ideologies to them. \citet{rela1} calculated the similarity between users according to interactive elements such as user mentions and retweets to classify the positions of new users and uses the relationship between users to optimize content features.

\paragraph{Semi-supervised and Unsupervised Methods.}
A full set of labeled users is not required for semi-supervised and unsupervised stance detection methods. The idea of language homogeneity between similar groups of people makes the user's relationship network important for label classification. Label propagation is widely used in the stance detection of users and tweets, which is based on the user following network~\cite{sesu1}, user retweet network~\cite{sesu2,sesu3} or tweet relationship~\cite{lp,lp2}, and then propagate and learn according to existing knowledge. The quality and imbalance of the initial label set will affect the effect of the model, which is also the current shortcoming of semi-supervised learning. STEM exploited interaction networks and used embeddings of similar and opposite stances to divide the speakers into stance-partitions~\cite{unsu4}. \citet{unsu5} utilized the target and the statement of the target to extract the effect triple to judge the target's position for arguments stance detection. In the task of user stance detection, \citep{unsu1} projected users into a common feature space for clustering by user retweets, hashtags, and other information, which had a good performance but also has a high computational cost. To make better use of the user relation network, Retweet-BERT~\cite{retweet-bert} trained similar users according to the idea of homogeneity and employ the large-scale training model, which achieved good results. At the same time, it emphasized the importance of graphs in social network research.

\subsection{Graph Neural Networks}
Graph neural networks have received growing attention recently~\cite{gnn1}. Many authors generalize deep learning networks (e.g., CNN, RNN) applicable to grids or sequences to arbitrarily structured graphs~\cite{gnn2,gnn3,gnn4,gcn,ggnn}. In their pioneering work, \citet{gcn} proposed a simplified graph neural network model, called graph convolutional networks (GCN), and achieved state-of-the-art classification results on several benchmark graph datasets. \citet{gcnso1} employed GCN on the social information graph which has two types of users and tweets to capture the documents’ social context. \citet{gcnso2} used GCN blocks to derive the target-adaptive graph representation of the context for stance detection. Graph Attention Network (GAT) introduced the multi-head attention mechanism to the GNN architecture and dynamically learn the weights on the edges when performing message passing~\cite{gat}. In addition, the graph neural networks with neighbor sampling and aggregation as the process~\cite{graphsage,pinsage} have achieved good development in terms of efficiency and effect. Some studies are dedicated to discovering different aggregation functions, so that information can be better disseminated and interacted with. SGC~\cite{sgc} and SSGC~\cite{ssgc} proposed different spectral graph convolutions to make calculations faster. GIN~~\cite{gin} was on the basis of the Weisfeiler-Lehman Test, modeling the injective multiset function for aggregation with strong representation ability. Noting that few supervised methods apply GNNs, we aim to propose a novel bipartite graph-based graph neural network for the supervised user stance detection task.

\section{Methodology}
In this section, we will describe our model in detail. First, we will introduce the dataset and data labeling method. Then we will explain the construction of the user-tweet bipartite graph. Finally, we will show how the model learns the representation of nodes by a homogeneous-heterogeneous joint information aggregation mechanism.

\subsection{Dataset and Labeling}
In the user's stance detection task, existing studies are grounded on user information or the relationship network between users (e.g. the following network and retweet network). We note that existing research mostly utilizes only user-related information. Therefore, aiming to analyze the user's position more comprehensively, we collect tweets between October 1, 2020 and November 2, 2020 by Twitter Streaming API. Then we filter the tweets related to the two candidates in the election: Joe Biden and Donald Trump by the following keywords: \textit{biden} or \textit{trump}. Finally, we get a total of 138.9 million tweets in English, along with the metadata (the time of the tweet, the author, etc.) and related user information (user ID, profile, location, etc.). 

For such an enormous amount of data, it is difficult to manually label all of them. We notice that the hashtags contained in a tweet are an explicit statement of position and are applied to detect user stances in previous studies~\cite{hashtag,label_bovet}. First, we obtain the hashtags with high frequency in the data, and three experts label the hashtags that plainly express the position. We interpret opposition to one candidate as support for the other. Then we expand the hashtag set through the correlation between hashtags and label similar hashtags as the same tag. As a result, the expanded hashtag set of 282 hashtags is used to label tweets. We select tweets that not only contain at least one hashtag but all hashtags in the tweet support the same candidate and then label the tweets according to the supported candidates.

User data is difficult to be labeled by user profile and personal information. First, 18.01\% of users have no profile, and their personal information (username, location, etc.) cannot express their position precisely. Secondly, many users with profiles do not take a stand in their profiles. However,  tweets and retweets of a user are an expression of the user's view. Therefore, we exploit their tweets and retweets to label the stance of users. Take user $u$, the set of tweets and retweets is $T_u = \{t_1^u, t_2^u,\cdots. t_n^u\}$, and the set of tweet label is $L_t = \{L_{t_1}, L_{t_2},\cdots, L_{t_n}\}$. The relative frequency $f_u$ of tweet positive examples of user $u$ is calculated as:
\begin{equation}
f_{u}=\frac{|\{L|L \in L_t\,\&\,L = 1\}|}{|L_t|}.
\end{equation}

\begin{figure}[!htb]
    \centering
    \includegraphics[width=0.47\textwidth]{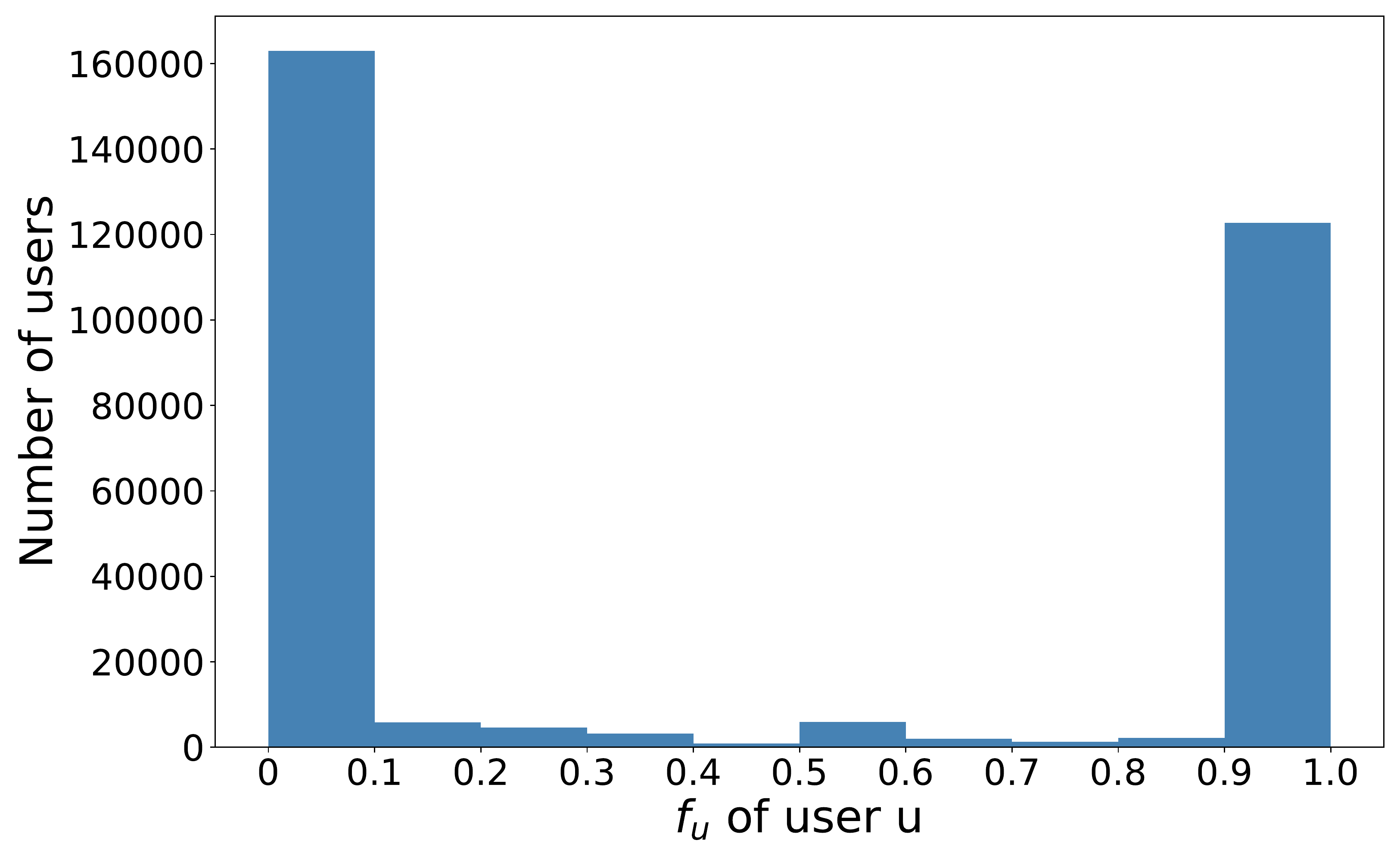}
    \caption{The distribution of $f_u$. The horizontal axis represents the interval of the relative frequency $f_u$ of the tweet's positive examples of user $u$. The vertical axis is the number of users whose $f_u$ is in a certain interval.}
    \label{fig:label_distuibution}
\end{figure}

We count the frequency distribution $D(f_u)$ of $f_u$ for all users. The frequency distribution histogram is shown in figure~\ref{fig:label_distuibution}
. From the figure, we can know that $D(f_u)$ roughly conforms to a U-shaped distribution, and there are many users with $f_u$ values of 0 and 1, which means that most users’ related tweets are of the same label. For other users, we need to annotate the label of each user by a threshold. Therefore, the label $L_u$ of user $u$ is:
\begin{equation}
L_{t}=
\begin{cases}
1& f_u \geq k_u\\
0& f_u \leq k_u
\end{cases}
\end{equation}
where $k_u$ is the threshold for user labels according to $f_u$. We set the value of $k_u$ to 0.5 based on $D(f_u)$ and get 177415 users who support Joe Biden and 134203 users who hold Donald Trump. We remove duplication in tweets and users and exclude hashtags used for annotation from all text. As a result, all labeled users and associated tweets make up our dataset. The statistics for the dataset are shown in Table~\ref{tab:data_info}. 

\begin{table}[!tb]
\centering
\begin{tabular}{ccccc}
    \toprule
    $N_{u}$ & $N_{t}$ & $R_{re}$ & Avg. $L_{t}$ & Avg. $W_{t}$\\
    \midrule
    311,618 & 1,123,749 & 0.57 & 159.09 & 24.21\\
    \bottomrule
\end{tabular}
\caption{The static information of the dataset. $N_{u}$ is the number of users. $N_{t}$ is the number of tweets. $R_{re}$ is the proportion of the number of `retweets` to all tweets. $L_{t}$ represents the length of tweet. $W_{t}$ indicates the number of words in a tweet.}
\label{tab:data_info}
\end{table}

\subsection{Bipartite Graph Construction}
Existing studies on user stance detection are grounded on the assumption of linguistic homogeneity among people with similar stances~\cite{homoid, homoid2}. Following graphs, retweet graphs, etc. are according to the relationship between users (following, reposting, etc.) and are widely used for stance detection. Users are the only node type in the graph whose profiles are utilized to learn user representations. The stances of users are detected based on known user tags, user personal information, and user relation network information. Therefore, users with richer information and more relation types can help construct more informative graphs. However, in real-world data, many users do not have detailed profiles, and the profiles have no clear meaning, which makes it difficult to detect the stances accurately only by the user's personal information. In this paper, we utilize user-related tweets and user information for user stance detection.

\begin{figure}[!htb]
    \centering
    \includegraphics[width=0.47\textwidth]{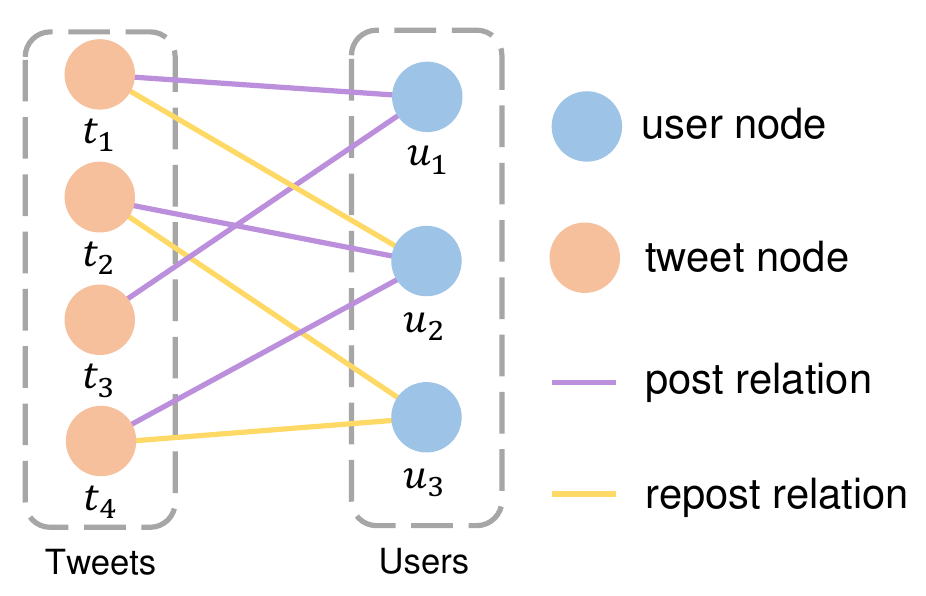}
    \caption{An example of the user-tweet bipartite graph.}
    \label{fig:utgraph}
\end{figure}

Taking user set $U$ and tweet set $T$, we get the post and retweet relation between users and tweets. $E_p = \{<u, t>, <t, u>|u \in U, t \in T, u\, post\,t\}$ is the directed post relation edges between users and tweets, $E_r = \{<u, t>, <t, u>|u \in U, t \in T, u\, repost\,t\}$ represents the directed retweet relation edges between users and tweets. The user-tweet directed bipartite graph $G=(V, E)$ is constructed by:
\begin{gather}
V = U \cup T,\\
E = E_p \cup E_r,
\end{gather}
where $V$ and $E$ are the node set and edge set of the graph respectively. As shown in Figure~\ref{fig:utgraph}, all user nodes are only connected to tweet nodes, so tweet nodes, specifically retweet nodes, serve as a bridge to connect users to users. We materialize the idea of ideological homogeneity as a directed bipartite graph, i.e., if a user retweets another user's tweet, the stances of the two users are similar. Compared with the user's personal information, tweets contain more information, and the position expressed in the text is more clear. Therefore, we add tweet information to the graph while retaining user relationships, which enables the model to learn better user node representations by using tweet information.

\subsection{DoubleH}

\paragraph{Preprocessing.}
For user data, we process scattered information of users by some templates. For example, for location information, we add `My location is` before the user's location information, and then splice this sentence after the profile of the user. Various information about the user can be integrated into a complete user profile by adding similar templates. For tweet data, we remove invalid information such as links and emoticons in the text. Since we label the data by hashtags, all relevant hashtags are removed from the user and tweet data. Sentence Transformers (S-BERT)~\cite{sbert} is a Siamese network optimized for sentence-level embeddings, which outperforms naive transformer-based methods for sentence-based tasks. In addition, it can massively reduce the time complexity. Therefore, we utilize the pre-trained model of S-BERT to directly retrieve the integrated user profiles and tweets and exploit the output of S-BERT as the initial representation of users and tweets.

\paragraph{Representation Learning.}

\renewcommand{\thealgorithm}{\arabic{algorithm}}
\begin{algorithm}[!htb]
\caption {DoubleH mini-batch forward propagation algorithm.}
\label{alg:mpm_algorithm}
\textbf{Input}: Graph $\mathcal{G(V, E)}$; Input features $\{x_v, \forall v \in \mathcal{B}\}$; \\
Depth $K$; Weight matrices $W^k, W_E^k$, $\forall k \in \{1,\cdots,K\}$; \\ 
Non-linearity $\sigma$; Layernorm $\mathcal{L}$; Dropout $\mathcal{D}$;
Neighbour sampler functions $\mathcal{N}_1 : u \rightarrow 2^u$ and $\mathcal{N}_2 : v \rightarrow 2^v$; \\
Aggregate functions $AGGREGATE_k$, $\forall k \in \{1,\cdots,K\}$.\\
\textbf{Output}: Representations $z_v$ for all $v \in \mathcal{B}$. \\
\begin{algorithmic}[1]
\STATE $\mathcal{B}^k \tiny{\leftarrow} \mathcal{B}$; \\
\FOR{$k = K,\cdots,1$}
\STATE $\mathcal{B}^{k-1} \tiny{\leftarrow} \mathcal{B}^k$; \\
\FOR{$u \in \mathcal{B}^k$}
\STATE $\mathcal{B}^{k-1} \tiny{\leftarrow} \mathcal{B}^{k-1} \cup \mathcal{N}_1(u)$; \\
\FOR{$v \in \mathcal{N}_1(u)$}
\STATE $\mathcal{B}^{k-1} \tiny{\leftarrow} \mathcal{B}^{k-1} \cup \mathcal{N}_2(v)$; \\ 
\ENDFOR
\ENDFOR
\ENDFOR

\STATE $h^0_u \leftarrow x_v$, $\forall v \in \mathcal{B}^0$; \\
\FOR{$k = 1,\cdots,K$}
\FOR{$u \in \mathcal{B}^k$}
\STATE $h^k_{\mathcal{N}_1} \tiny{\leftarrow} \sigma (W^k_{E(u)} \mathcal{D}(\mathcal{L}(h^{k-1}_{u^{\prime}}, \forall u^{\prime} \in \mathcal{N}_1(u)))) $; \\
\STATE $h^k_{\mathcal{N}_2} \tiny{\leftarrow} \sigma (W^k_{E(u)} \mathcal{D}(\mathcal{L}(h^{k-1}_{u^{\prime}}, \forall u^{\prime} \in \mathcal{N}_2(\mathcal{N}_1(u))))) $; 
\STATE $h^k_{\mathcal{N}_(u)} \tiny{\leftarrow} AGGREGATE_k(h^k_{\mathcal{N}_1} \cup h^k_{\mathcal{N}_2})$; 
\STATE $h^k_u \tiny{\leftarrow} \sigma (W^k \cdot CONCAT(h^{k-1}_u, h^k_{\mathcal{N}_(u)})$;
\STATE $h^k_u \tiny{\leftarrow} h^k_u/\vert\vert h^k_u \vert\vert _2$;
\ENDFOR
\ENDFOR
\end{algorithmic}
\end{algorithm}

According to the user-tweet bipartite graph, we aim to learn the user node representation with user and tweet information. Specifically, following the message passing mechanism in GNNs, we intend to iteratively update the node representation of the bipartite graph with both user and tweet nodes at each layer. In the bipartite graph, the one-hop neighbors of each node are different node types from the center node, and the two-hop neighbors are of the same node type as the center node. For user nodes, the information of other users is homogeneous, while the information of tweets is heterogeneous. In our model, we consider homogeneous and heterogeneous information separately. Algorithm~\ref{alg:mpm_algorithm} describes the update process of each node embedding.

Since we exploit the homogeneous and heterogeneous information of each node, in each iteration, we need to collect the one-hop and two-hop neighbors of the node. In order to improve the training efficiency, we use the mini-batch and the method of neighbor sampling to generate neighbors. First, for each node in the batch, we sample its one-hop neighbors by $\mathcal{N}_1$, and then another neighbor sampler $\mathcal{N}_2$ samples the neighbors of the one-hop neighbors. $\mathcal{N}_1(v) $and $\mathcal{N}_2(v)$ are both fixed-size, uniform draw from the set $\{u \in \mathcal{V} : (u, v) \in \mathcal{E}\}$. 

For user nodes, all one-hop neighbors are tweet nodes, and all two-hop neighbors are user nodes. Tweets are direct expressions of the user's position, so tweet information can help update the user's node representation. In addition, the two-hop neighbors of user nodes provide network information on retweet relationships between users. In other words, homogeneous information provides network relationships, and heterogeneous information adds additional information to nodes. Therefore, we intend to treat these two types of information separately. First, we process heterogeneous information $\{h^{k-1}_{u^{\prime}}, \forall u^{\prime} \in \mathcal{N}_1(u)\}$ into $h^k_{\mathcal{N}_1}$ by layernorm $\mathcal{L}$, linear weight matrices $W^k_E$ and activation function $\sigma$. Similarly, homogeneous information $\{h^{k-1}_{u^{\prime}}, \forall u^{\prime} \in \mathcal{N}_2(\mathcal{N}_1(u))\} $ is processed into $h^k_{\mathcal{N}_2}$. To accurately utilize different types of edges in the bipartite graph, we use a type linear layer $W^k_E$. In each iteration, two $W^k_{E(u)}$ in line 14 and line 15 of the algorithm ~\ref{alg:mpm_algorithm} are not the same for different node types. For heterogeneous neighbor information $h^k_{\mathcal{N}_1}$ and homogeneous neighbor information $h^k_{\mathcal{N}_2}$, we aggregate them into a neighborhood vector $h^k_{\mathcal{N}(u)}$ by aggregation function. We concatenate the node's previous layer representation $h^{k-1}_u$ with the aggregated neighborhood vector $h^k_{\mathcal{N}(u)}$. The representation of nodes is finally updated by the linear layer, nonlinear layer, and normalization layer.

After $K$ iterations, the model updates the representations of all nodes. The predicted label of the original user is computed as:
\begin{equation}
y_u = softmax(Wh^K_u + b),
\end{equation}
where $h^K_u$ is the representation vector of nodes of the final layer, and $W$ and $b$ are the weight and bias respectively. The goal of training is to minimize the cross-entropy between the ground truth label and the predicted label:
\begin{equation}
loss = -\sum_{u} g_u\log(y_u),
\end{equation}
where $g_u$ is the one-hot vector of the ground truth label.

\begin{table*}[!ht]
\centering
\begin{tabular}{llccccc}
    \toprule
    Model Type & Model & User Info. & Tweet Info. & Acc. & Auc & F1\\
    \midrule
    \multirow{6}{*}{Homogeneity Models} 
    & GCN & \Checkmark & \Checkmark & 76.41 & 75.32 & 71.12 \\
    & GAT & \Checkmark & \Checkmark & 77.47 & 76.60 & 72.90 \\
    & PinSage & \Checkmark & \Checkmark & 72.92 & 71.26 & 68.21 \\ 
    & GraphSAGE & \Checkmark & \Checkmark & 84.48 & 83.94 & 81.62 \\
    & GIN & \Checkmark & \Checkmark & 83.52 & 83.00 & 80.55 \\
    & Retweet-BERT & \Checkmark & \XSolidBrush & 68.31 & 65.71 & 56.12 \\ 
    & GPPT & \Checkmark & \Checkmark & 81.77 & 80.71 & 77.98 \\ 
    \midrule
    \multirow{2}{*}{Heterogeneous Models} 
    & RGCN & \Checkmark & \Checkmark & 76.40 & 75.49 & 71.55\\
    & RGAT & \Checkmark & \Checkmark & 77.57 & 76.86 & 73.47 \\
    \midrule
    \textbf{DoubleH(our model)} & \textbf{DoubleH} & \Checkmark & \Checkmark & \textbf{85.67} & \textbf{85.38} & \textbf{83.36} \\ 
    \bottomrule
\end{tabular}
\caption{Performance of different models on the dataset. Bold indicates the top-performing method.}
\label{tab:all_result}
\end{table*}

\section{Experiments}
In this section, we evaluate the effectiveness of our model\footnote{The code will be available after the acceptance of this paper.} and report the experimental results.

\paragraph{Baselines.}
In this paper, we focus on addressing the utilization of user-tweet bipartite graphs in user stance detection, so we mainly choose the graph-based methods as our baselines. We divide the baseline models into two categories: (i) Homogeneous graph methods, including GCN~\cite{gcn}, GAT~\cite{gat}, GraphSAGE~\cite{graphsage}, PinSage~\cite{pinsage}, GIN~\cite{gin}, Retweet-BERT~\cite{retweet-bert} and GPPT~\cite{gppt}. Retweet-BERT is a graph neural network method for user ideology detection. It employs the user's retweet graph and user profile information, combined with the large-scale pre-trained model to classify the ideology of users. GPPT is a recently proposed graph neural network model based on pre-training and prompt Tuning. The model reduces the training objective gap of the model by pre-training, prompt tuning functions, and fine-tuning. It has achieved the most advanced performance on graph-related tasks. (ii) Heterogeneous graph method: including RGAT~\cite{rgat} and RGCN~\cite{rgcn}. 

\paragraph{Setup.}
We randomly divide the training set into the training set and the validation set at a ratio of 9:1. We use the Adam optimizer with an initial learning rate of 0.001 and set the dropout rate to 0.1. The hidden size of the model is 768, and the batch size is 1024. The layer of the model is between 1 and 3. We set the sample size of the neighbor sampling function $\mathcal{N}_1$ to 10, and the sample size of the neighbor sampling function $\mathcal{N}_2$ is also 10. The aggregate functions are set to sum in our model. For DoubleH and all baselines, we use the BERT$_{base}$ model as the pre-training model for S-BERT. 

\subsection{Experimental Results}
Table~\ref{tab:all_result} presents the performance of DoubleH and baselines, from which we can observe that the performance of DoubleH is generally better than other baselines. According to the homogeneous graph approach, user and tweet nodes are treated as the same node type. GraphSAGE achieves the best performance, second only to DoubleH in all three metrics. In addition, GIN and GPPT have also achieved competitive performance. The performances of GCN and GAT are mediocre. Pinsage is a graph neural network based on a bipartite graph, which is mainly used in recommendation systems. Pinsage finds the important neighbors of the target node by random walk and aggregates these important neighbor nodes during information aggregation. In the recommendation system, the weight gap between users and items is obvious, so it is relevant for nodes to find important neighbors. However, in our dataset, each tweet of a user is an expression of a position, and tweets with low side weights may have more obvious position bias. Therefore, the performance of Pinsage is not as prominent as other methods. Retweet-BERT employs the retweet network of users to reduce the similarity between neighboring users as the goal, and the pre-training is grounded on the profiles of users. The user's ideology is obtained by the pre-trained representation. We extract the retweet subnetwork of the data and input it into the Retweet-BERT model for pre-training. There are two reasons for the deficient performance of the model. First, the model is based on the idea of linguistic homogeneity and fully utilizes the user's network relationship. But in the real world, it is not only the users who retweet who have a position. Many users do not retweet tweets, and their tweets are not retweeted by other users, which leads that more than half of the users do not appear in the retweet network and Retweet-BERT not pre-training with all users. Second, as we mentioned in Section 4.1, some users do not have a profile in our data, and the data quality of the profile of the remaining users is uneven. Therefore, it is not enough to detect the user's stance only by the user's information, and user-related tweets can greatly improve the performance of the model on the user's stance detection task.  For heterogeneous graph methods, RGCN and RGAT perform comparably to GCN and GAT respectively. Heterogeneous graph neural network models utilize heterogeneous attribute information of different nodes and edges to update node representations. For the user-tweet bipartite graph, although user nodes and tweet nodes are different types of nodes, they are both texts, and there is no obvious difference in attributes. Therefore, better utilization of different relationships between nodes is the main reason for the improved performance.

In other graph neural network models, each node interacts with one-hop neighbors directly, and the information of two-hop neighbors passes the message to the target node through one-hop neighbors in the next layer. In the user-tweet bipartite graph, the interaction between a node and its one-hop neighbor represents the information interaction between the user and tweet, and the interaction between the node and the two-hop neighbor represents the information interaction between the user and other users. DoubleH enables each user node to directly obtain user network information by allowing nodes to straightly interact with two-hop neighbors, which makes the node obtain more accurate information than other models. DoubleH outperforms other baseline models on all metrics, achieving state-of-the-art performance. Next, we will further analyze the ablation of information utilization, the layer of the models, and the efficiency of DoubleH.

\subsection{Ablation Study}
DoubleH achieves good performance by effectively aggregating homogeneous and heterogeneous information. To further understand the role of these two types of information, we performed ablation learning on one-layer and two-layer DoubleH with only one of the information for aggregation each time. The results are shown in Table~\ref{tab:ablation}. 

\begin{table}[!ht]
\centering
  \begin{tabular}{lccc}
    \toprule
    Model & Hete Info. & Homo Info. & F1\\
    \midrule
    \multirow{3}{*}{1-Layer DoubleH} 
    & \XSolidBrush & \Checkmark & 67.81 \\
    & \Checkmark & \XSolidBrush & 81.80 \\
    & \Checkmark & \Checkmark & 82.77 \\
    \midrule
    \multirow{3}{*}{2-Layer DoubleH} 
    & \XSolidBrush & \Checkmark & 69.02 \\
    & \Checkmark & \XSolidBrush & 82.45 \\
    & \Checkmark & \Checkmark & 83.36 \\
    \bottomrule
  \end{tabular}
\caption{Ablation learning of heterogeneous and homogeneous information on DoubleH with 1 and 2 layers.}
\label{tab:ablation}
\end{table}

The results indicate that the performance of the model will not degrade too much when only heterogeneous information is exploited in the aggregation, the performance of the model will drop significantly when using only homogeneous information. For homogeneous information, if only homogeneous information is utilized for each aggregation, the model directly ignores its one-hop neighbors, and user nodes only interact with user nodes. Therefore, only retweet network information is exploited in aggregation, causing DoubleH to behave similarly to Retweet-BERT. On the other hand, heterogeneous information exhibits a more important role. The results in Table~\ref{tab:ablation} show that 1-layer and 2-layer DoubleH perform comparably to GraphSAGE and GIN when only using heterogeneous information. Among the existing graph neural network models, messages are transmitted via direct neighbors, and nodes can only interact with their one-hop neighbors at a time. DoubleH extracts and aggregates homogeneous information and heterogeneous information respectively. The 1-hop and 2-hop neighbors become the `new one-hop neighbors` of the central node, which breaks the traditional node interaction mechanism. For the user stance detection task, this new mechanism can help balance and utilize the tweet information and network information of the bipartite graph, so that model can get a better node representation. In addition, with the emergence of more diverse graph data structures, this mechanism can also be generalized to interact with more hops.

\subsection{Model Layer Number}
In the graph neural network methods, the number of layers k of the model represents that each node can fuse k-hop neighbor information. Table~\ref{tab:layer_result} shows the performance of some baseline models with different layer numbers. 

\begin{table}[!ht]
\centering
  \begin{tabular}{lccc}
    \toprule
    Model & Layer Number & Acc. & F1\\
    \midrule
    \multirow{2}{*}{GCN} 
    & 1 & \underline{76.41} & \underline{71.12} \\
    & 2 & 74.47 & 69.73 \\
    \hline
    \multirow{2}{*}{GAT} 
    & 1 & \underline{77.47} & \underline{72.90} \\
    & 2 & 74.54 & 70.82 \\
    \hline
    \multirow{2}{*}{RGCN} 
    & 1 & \underline{76.40} & \underline{71.55} \\
    & 2 & 74.12 & 69.35 \\
    \hline
    \multirow{2}{*}{RGAT} 
    & 1 &\underline{77.57} & \underline{73.47} \\
    & 2 & 76.10 & 72.86 \\
    \hline
    \multirow{3}{*}{GraphSAGE} 
    & 1 & 76.89 & 72.24 \\
    & 2 & 84.31 & 81.46 \\
    & 2 & \underline{84.48} & \underline{81.62} \\
    \hline
    \multirow{3}{*}{GIN} 
    & 1 & 74.70 & 69.20 \\
    & 2 & \underline{83.52} & \underline{80.55} \\
    & 3 & 83.44 & 80.53 \\
    \hline
    \multirow{3}{*}{DoubleH} 
    & 1 & 85.23 & 82.77 \\
    & 2 & 85.67 & \underline{\textbf{83.36}} \\
    & 3 & \underline{\textbf{85.84}} & 83.24 \\
    \bottomrule
  \end{tabular}
\caption{Performance of models with different layers. Bold indicates the best performance, and underlined text refers to the best performance of the same model at different layers.}
\label{tab:layer_result}
\end{table}

We can observe that GCN, GAT, RGCN, and RGAT degrade the performance when the number of layers is increased to 2. The performance of the 1-layer GIN model is slightly inferior to GCN and GAT, while GIN with 2 layers shows stable performance. The performance of the GraphSAGE model improves as the number of layers increases. The 3-layer GraphSAGE model is the best in all baselines, which is comparable to the performance of the 2-layer model. From the performance of the baseline models, the 1-layer models do not perform well, because each user node only integrates the heterogeneous information of its direct neighbors, and loses the homogeneous information of its neighbors from the second hop. When the number of layers of the model increases, the stronger models show the ability to fuse multi-hop information. DoubleH allows nodes to directly interact with one-hop neighbors and two-hop neighbors when fusing information, so DoubleH with one layer outperforms all baseline models. The 2-layer and 3-layer DoubleH show better performance. 

\subsection{Efficiency Analysis}

\begin{figure}[!htb]
    \centering
    \includegraphics[width=0.45\textwidth]{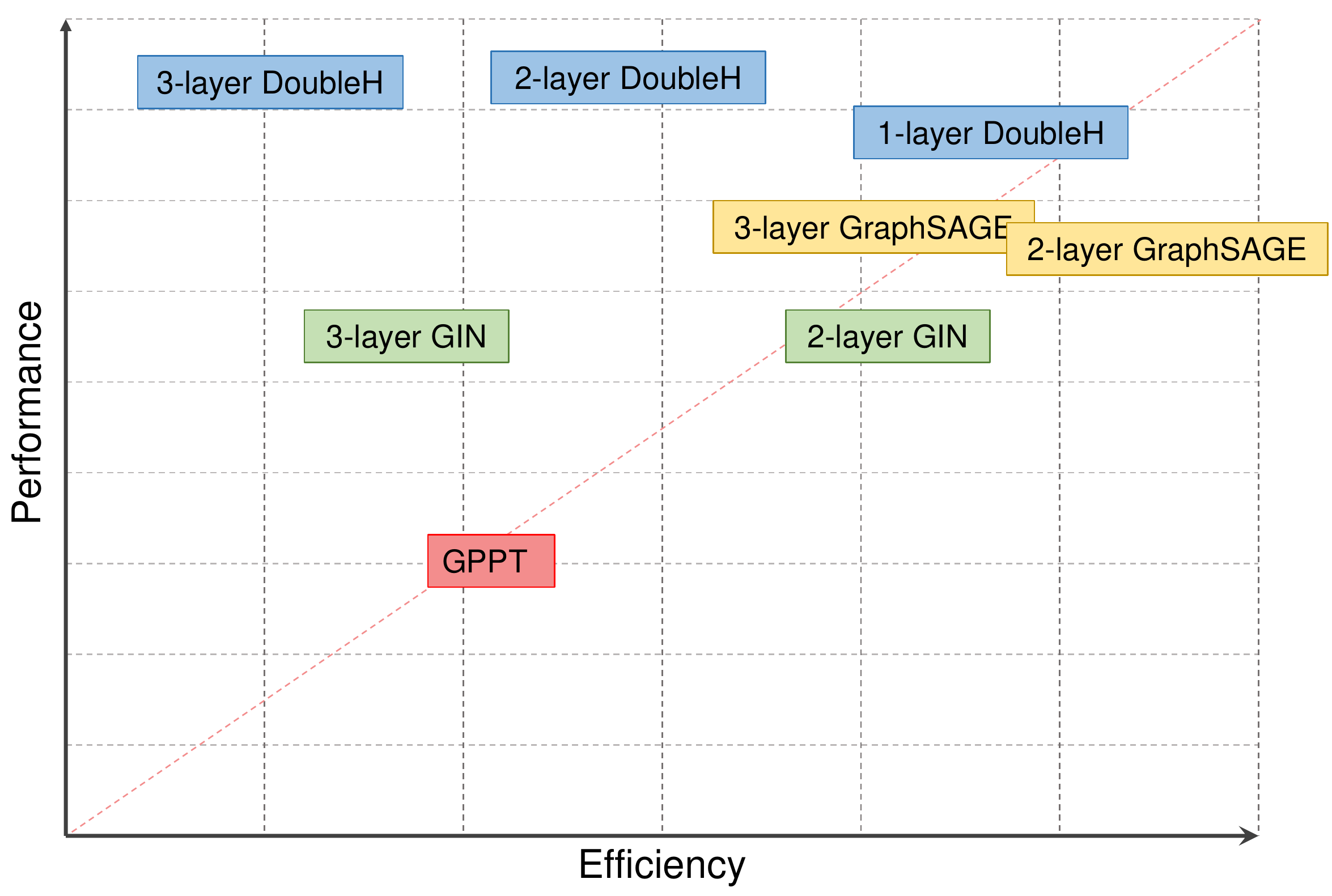}
    \caption{The efficiency-performance graph of baselines. The horizontal axis is the model, and the vertical axis is the performance of the model based on the F1 value.}
    \label{fig:efficiency}
\end{figure}

Each additional layer of the model will increase the computational complexity of the model, thereby increasing the training time of the model. Among our baselines, the Retweet-BERT model is inductive, and the other models are transductive. The transductive models can achieve better performance, but need to retrain when encountering new incoming data in practical applications. For the stance detection task of social media users, the accuracy rate of the model, F1 value, and other metrics have received too much attention. However, the model should serve practical problems, and the best performance model with a large amount of calculation is difficult to be applied. Therefore, the efficiency of the model deserves to be taken into consideration. We take the training time as the efficiency of the model. The shorter the training time, the higher the efficiency of the model. We record the efficiency of the model with the F1 value above 0.75, Figure~\ref{fig:efficiency} shows the efficiency-performance graph for different baselines. From the figure, we noticed that the 2-layer DoubleH has the best performance with lower efficiency than GraphSAGE and GIN. The overall performance of GraphSAGE is better than that of GIN. The 3-layer DoubleH with the second-best F1 value needs the longest training time among all models, so it is not recommended to be employed for practical problems. The 1-layer DoubleH has both good performance and efficiency, becoming the model with the best overall performance.

\section{Conclusion}
In this paper, we propose a novel model based on the user-tweet bipartite graph for user stance detection. We first collect social media data about the 2020 US presidential election. Then we annotate the tweets based on manually tagged hashtags and label the user's stance according to the label ratio of related tweets. Our model extracts and exploits both homogeneous and heterogeneous information of each node to learn node representations. DoubleH enables each node to interact directly with its one-hop and two-hop neighbors at the structural level and handles homogeneous information and heterogeneous information separately at the information level. Experimental results demonstrate the ability of DoubleH to utilize both homogeneous and heterogeneous information. Our model can achieve state-of-the-art performance while achieving both high efficiency and good performance by reducing the number of layers. 

On the other hand, DoubleH performs information interaction on the entire user-tweet bipartite graph. For the stance detection of new users, the model needs to rebuild the graph and retrain, making the model unfriendly to new incoming data, which is also the limitation of all transductive models. We will consider this issue in future work and improve our model.


\end{document}